\documentclass[10pt]{iopart}
\usepackage{graphicx}
\usepackage{amsmath}
\usepackage{caption}
\usepackage{multirow}
\usepackage[table,xcdraw]{xcolor}
\usepackage{subcaption}
\begin{document}

\title[Measurement of neutron total cross-section of Mg and  MgF\textsubscript{2}  with fine multi-scattering corrections.]{Measurement of neutron total cross-section of Mg and MgF\textsubscript{2} at MONNET (JRC-Geel) from transmission with fine multi-scattering corrections}

\author{Marco Antonio Martínez-Cañadas$^1$, Pablo Torres-Sánchez$^{1,2,*}$, Miguel Macías$^3$, Javier Praena$^1$, Cristiano L. Fontana$^3$, Cédric Bonaldi$^3$, Wouter Geerts$^3$, Stephan Oberstedt$^3$ and Ignacio Porras$^1$}
\address{$^1$ Universidad de Granada, Av. de Fuente Nueva, s/n, Beiro, 18071 Granada, Spain}
\address{$^2$ Instituto de Física Corpuscular, (CSIC-UV), Carrer del Catedrátic José Beltrán Martinez, 2, 46980 Paterna, Valencia, Spain}
\address{$^3$ European Commission, Joint Research Centre (JRC), Retieseweg 111, 2440 Geel, Belgium}
\ead{Pablo.Torres@ific.uv.es}

\begin{abstract}
Neutron-transmission measurements through samples of magnesium fluoride (MgF$_2$) and pure magnesium were performed to obtain the (n, tot) cross section for all isotopes involved, $^{19}$F and $^{24-26}$Mg. Lithium-glass detectors were used in conjunction with the neutron time-of-flight technique. The measurement campaign was performed at the MONNET fast-neutron source of the European Commission Joint Research Centre (JRC-Geel, Belgium). Highly precise corrections for multiple scattering were calculated using a sophisticated iterative method based on Monte Carlo simulations with the MCNP6.3 code, accounting for the effects of the experimental setup. With the SAMMY code, an R-Matrix analysis of the experimental data was performed. The extracted cross-sections, resonance spin and parity as well as the limitations of the method are carefully discussed.
\end{abstract}

\vspace{2pc}

\submitto{\jpg}
\maketitle
\ioptwocol

\section{Introduction}
\subsection{Applications}\label{sec:applications}
The measurement of the cross-sections of \textsuperscript{19}F(n, tot), \textsuperscript{24}Mg(n, tot), \textsuperscript{25}Mg(n, tot) and \textsuperscript{26}Mg(n, tot) reactions is very valuable in medical physics, more particularly for Boron Neutron Capture Therapy (BNCT). 

In BNCT, the use of reliable high-intensity neutron beams is of paramount importance \cite{Kreiner2016}. The MgF\textsubscript{2} material is of great relevance for the accelerator-based production of neutron beams for BNCT, as extensively proven by several proposals that consider using Mg and F as part of the moderating core in their Beam Shaping Assembly (BSA) designs \cite{TRO4713, daico2025, ZHU20241813}, including the NeMeSiS project from the University of Granada \cite{Porras2020BNCT}. Previous works have explored the effects of thermal-neutron scattering in MgF\textsubscript{2} \cite{HU20231280}. Here, our goal is to extend that analysis to the epithermal range, the most relevant one in the moderation process of neutrons in BNCT. To that end, a BSA model was designed \cite{TorresSanchez2021}, that features a moderating core of MgF\textsubscript{2}. This has led to a patent by our group \cite{PCTES2021070607}.

Taking advantage of the combination of resonances of Mg and F isotopes, the moderation of the high-energy neutrons is enhanced. When in conjunction with other materials, Monte Carlo simulations showed that it could effectively reduce the maximum energy to 200 keV when using \textsuperscript{7}Li(p,n) with a 2.1~MeV neutron source \cite{TorresSanchez2021}. A series of measurements is required to disentangle the nuclear-data related uncertainties, and a precise measurement of the resonances of interest is the first step. 

In astrophysics, \textsuperscript{25}Mg and \textsuperscript{26}Mg are very important neutron poisons via neutron capture, competing with capture on \textsuperscript{56}Fe. The reaction \textsuperscript{22}Ne($\alpha$,n)\textsuperscript{25}Mg is one of the most important neutron sources in Red Giant stars, with a very uncertain reaction rate. Understanding the properties of the \textsuperscript{26}Mg resonances is the key to improving the issue, so any information on the width of any of its resonances is of interest. General knowledge of the cross-sections obtained in this work is also useful for Maxwellian Averaged Cross-Section calculations of astrophysical interest.
Regarding fluorine, it is particularly interesting in the field of fusion reactors. Tritium breeding blankets are a key component in fusion reactor designs, using Li for the task. Due to its physical properties, pure lithium is not a suitable option and therefore compounds containing it are used \cite{HERNANDEZ2018243, YANG1981585}. Some of those compounds contain fluorine, such as LiF \cite{TERAI199897} or the salts Li\textsubscript{2}BeF\textsubscript{4} (FLiBe) and LiF–NaF–BeF2 (FLiNaBe) \cite{DELPECH201034}. FLiBe is specially interesting for tritium breeding due to its synergies with its other constituents (neutron multiplication through \textsuperscript{9}Be(n, 2n) reactions, and tritium production from isotope \textsuperscript{7}Li  via (n, n'$\alpha$) reaction with no neutron consumption), as well as its properties as a coolant. Fluorine is also a critical element for molten salt reactors, in which the fuel is in a molten salt solution that at the same time serves as coolant, like LiF-BeF\textsubscript{2}-UF\textsubscript{4}. The reaction \textsuperscript{19}F(n,n’) is also important in fusion systems (as well as in fission reactors) due to the gamma radiation of high energy produced from inelastic neutron scattering in fusion reactor components. Such radiation contributes to the deposited heat it also modifies the neutron energy spectrum (whose shape is crucial in tritium breeding). Different sensitivity studies analyze the impact on propagation of \textsuperscript{19}F(n,n’) uncertainty in fusion and fission designs that use fluorine salts \cite{sensibility1, sensibility2}.

Besides astrophysical, medical and energy production applications, the data reported in this work can also be very valuable towards the elaboration of future nuclear evaluations.

A key remark is that the cross section that we are trying to measure is a total cross section of a reaction dominated by elastic scattering, so our approach of dedicated analysis of the multi-scattering effects is particularly relevant. Our measured spectra will be corrected by multi-scattering effects, but this correction strongly depends as well on what is measured. The approach and corrections presented in this work can also be seen as a general result in itself: the description of a method useful for any cross-section measurement of this kind.

\subsection{Previous measurements}
The energy range scoped in this work covers a range from 10 keV to 360 keV. For such energy range, scarce experimental data currently exist for the total cross-section of magnesium isotopes, while \textsuperscript{19}F is better known. More particularly: 

\begin{itemize}
    \item For \textsuperscript{19}F, several measurements were carried out with a general agreement among them \cite{newson1961, hibdon1964, macklin1973, singh1974_F}.
    \item For \textsuperscript{24}Mg, there is only one measurement of the total cross-section below 1 MeV by A. J. Elwyn and R. O. Lane from 1962 \cite{elwyn1962}, and solely covers the range between 164 keV to 455 keV, excluding the strong resonance at 83 keV. There is also a lack of data on the elastic cross-section, with measurements only made above 3 MeV.
    \item For \textsuperscript{25}Mg, it was recently measured in 2017 with high resolution by transmission at GELINA by Massimi \textit{et al.} \cite{massimi2017}, though their data is not available in EXFOR.
    \item For \textsuperscript{26}Mg, a single measurement from 1959 by Newson \textit{et al.} \cite{newson1959} is recorded, with no agreement with last ENDF evaluation besides the 299 keV resonance. 
\end{itemize} 

The last ENDF evaluation (ENDF-VIII.1) \cite{ENDF_evaluation} uses resonance parameters from \cite{ATLAS_Mughabghab} for \textsuperscript{24}Mg below 520 keV, for \textsuperscript{25}Mg below 220 keV (but slightly modified to better match experimental results from \cite{hibdon1964} and \cite{singh1974_Mg}) and for \textsuperscript{26}Mg below 450 keV. As for \textsuperscript{19}F, results from SAMMY R-matrix fit of \cite{larson1976} are directly provided. Unconfirmed values of spin and parity for resonance at 97.95 keV are provided by \cite{ATLAS_Mughabghab}, but later works \cite{singh1974_F} suggest 1\textsuperscript{-} as the right one, which is also confirmed by our results. 

Similar analyses were performed by \cite{singh1974_F, singh1974_Mg} in 1974 for magnesium and teflon (CF\textsubscript{2})\textsubscript{n} samples. More realistic results are expected from this work thanks to the fine multi-scattering corrections of the method presented in following sections.

Finally, with respect to fluorine isotope, even though more data has been measured for the total cross-section, the first excited state of \textsuperscript{19}F is at 109.894 keV while the second one is at 197.143 keV, meaning that the channel (n,n') will contribute to the total cross-section measured in part of the energy range covered in our measurement. Four measurements are found in EXFOR of \textsuperscript{19}F(n,n') cross-section:

\begin{itemize}
    \item Freeman \textit{et al.} \cite{Freeman01051957} in 1957, who measured the partial channels separately.
    \item Broder \textit{et al.} \cite{Broder1969} in 1969.
    \item Roger \textit{et al.} \cite{Roger1974} in 1974.
    \item Lashuk \textit{et al.} \cite{Lashuk1994} in 1994, who also measured the partial cross-section of independent levels.
\end{itemize} 

\section{Methodology}
With the main goal of obtaining a new measurement of the MgF\textsubscript{2} neutron cross-section, an experiment was performed. All aspects regarding its set-up will be presented in this section: the experiment, the facility, the detectors used, the samples, and corrections due to experimental conditions.

\begin{figure}[]
\begin{center}
\includegraphics[width=0.47\textwidth]{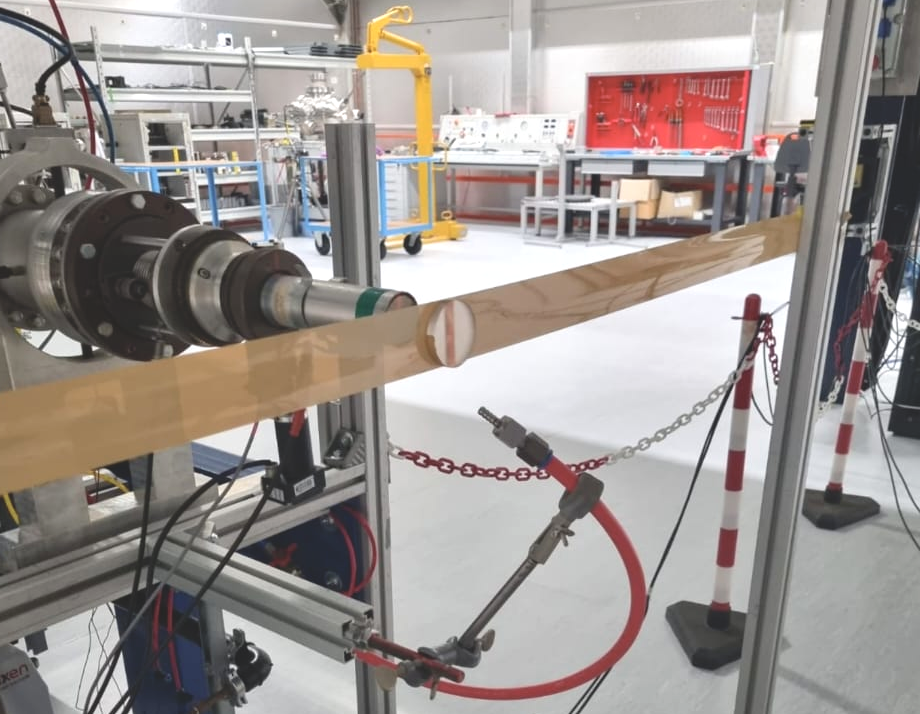}
\end{center}
\caption{\label{fig:experimental_picture} Photograph of the MgF\textsubscript{2} sample (thick) placed in front of the neutron source.}
\end{figure}

\subsection{The Monnet facility}
The experiment was performed at the Joint Research Centre in Geel, Belgium \cite{JRC_Tandem_Accelerator}. The MONNET facility offers a high-intensity fast neutron source, driven by a vertical 3.5~MV Tandem accelerator producing either continuous or pulsed beams of protons deuterons ions impinging on lithium, deuterium or tritium targets, and whose yield were studied in detail in \cite{MACIAS2024109304, Fontana2023}. The produced ion beams can reach currents of up to 50 $\mathrm{\mu}A$ on target in continuous mode and up to 10 $\mathrm{\mu}A$ in pulsed mode. Repetition rates for pulsed mode are 2.5, 1.25 or 0.625~MHz and the ion pulse-lengths are $<$ 2 ns FWHM, able to operates continuously, 24 hours a day, 7 days a week. Frequency of 0.625~MHz, acceleration voltage of $2.100\pm 0.001$~MeV, proton beam on an air-cooled lithium fluoride target. The experimental hall is a low-scattering environment with walls at least 10~m away from the neutron-producing target and the floor is constituted by light aluminum profiles 3~m above the solid ground (enough to support the beamline weight but minimized to diminish backscattering effects).

\subsection{Detectors and Data Acquisition System}
\begin{figure}[]
\includegraphics[width=0.5\textwidth]{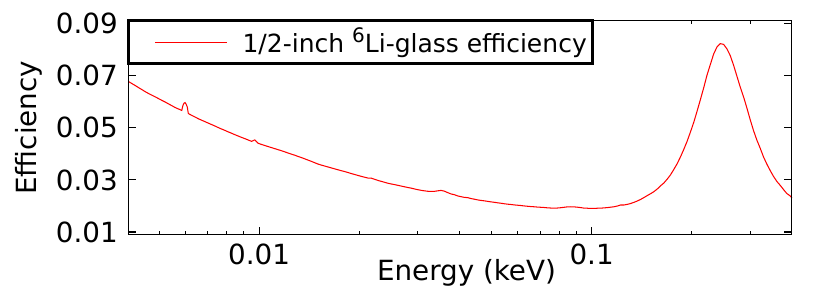}
\caption{\label{fig:6liglass_efficiency} Efficiency of the \textsuperscript{6}Li-glass detector in the energy range of interest, obtained with Monte Carlo simulation using MCNP6.3 code.}
\end{figure}

The Time of Flight (TOF) technique was used with a flight path of 2~m for the measurements in conjunction with several detectors. The main neutron detector was a 1/2-inch lithium glass detector (\textsuperscript{6}Li-glass). A detailed diagram of the used detector can be found in \cite{tesis_miguelmacias}, and its efficiency is shown on Figure \ref{fig:6liglass_efficiency}. Two extra detectors were used in the experimental set up for normalization: a cerium bromide detector (CeBr\textsubscript{3}) for gamma radiation detection 1~m away from the target and -30$^{\circ}$ with respect to the beam, and a DePangher neutron detector (Proportional Long Counter) 2~m away from target and at -57$^{\circ}$, fully characterized \cite{Thomas2018}. Resolution in TOF binning for all measurements is 2 ns.

The ABCD framework, maintained by the MONNET team \cite{ABCD_Documentation, fontana1, fontana2, fontana3, fontana4}, served as the front end for the data acquisition system in this experiment. ABCD is responsible for initializing the DAQ, reading the data from the hardware, storing, and processing the acquired waveforms. For our setup, we used the CAEN waveform digitizer model DT5730 \cite{CAEN_DT5730_DT5725_Manual}, a desktop module that operates as the DAQ. The DT5730 features eight channels with flash ADCs of 14 bits and 500 Msamples/s, making it suitable for high-resolution signal acquisition.

\subsection{Samples and source characterization}
The Mg sample was purchased from Goodfellow \cite{GoodfellowMaterials} and MgF\textsubscript{2} from \cite{RobsonScientific}, with nominal purities of at least 99.9\% according to the manufacturer. Three different samples were used (two of magnesium fluoride, one of magnesium metal) and their dimensions are shown in Table \ref{tab:dimensions}.

\begin{table}
\caption{Mass, thickness and diameter of the transmission samples.}\label{tab:dimensions}
\footnotesize
\centering
\resizebox{0.5\textwidth}{!}{\begin{tabular}{@{} cccc @{}}
\br
   Sample & Mass (mg) & Thickness (mm) & Diameter (mm) \\
\mr
MgF\textsubscript{2} thin  & 12536.2 $\pm$ 0.1  & 2.02 $\pm$ 0.01           & 50.00 $\pm$ 0.01         \\
MgF\textsubscript{2} thick & 54700.5 $\pm$ 0.1  & 8.84 $\pm$ 0.01           & 50.00 $\pm$ 0.01         \\
Pure Mg    & 19449.2 $\pm$ 0.1  & 5.80 $\pm$ 0.01           & 50.00 $\pm$ 0.01         \\ \hline
\br
\end{tabular}}
\end{table}

\begin{figure}
     \centering
     \begin{subfigure}[b]{0.46\textwidth}
         \centering
         \includegraphics[width=\textwidth]{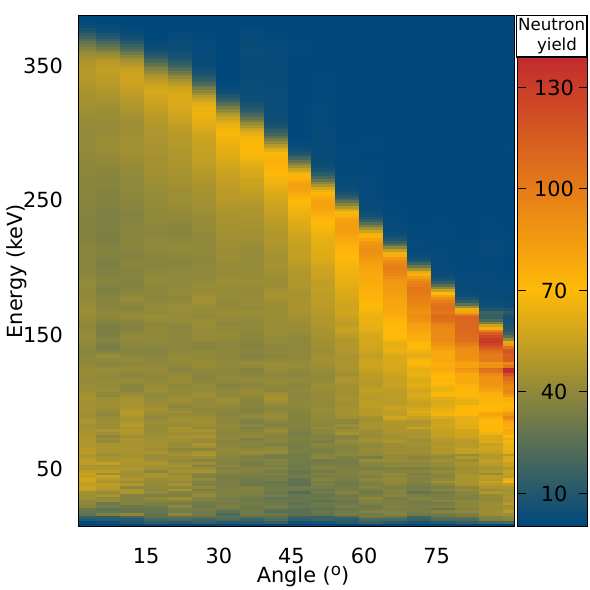}
        \caption{Values of the spectrum of neutrons emitted from the LiF target as a function of the energy and the angle.}
        \label{fig:values_2D_neutronspectra}
     \end{subfigure}
     \hfill
     \begin{subfigure}[b]{0.46\textwidth}
         \centering
         \includegraphics[width=\textwidth]{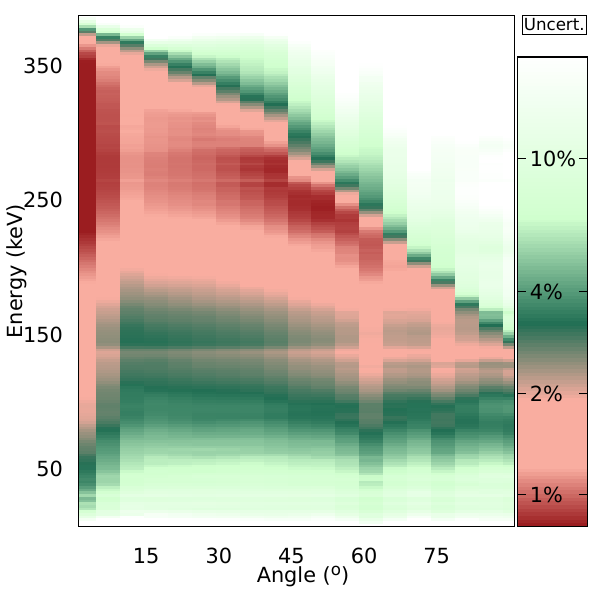}
        \caption{Uncertainties of the spectrum of neutrons emitted from the LiF target as a function of the energy and the angle.}
        \label{fig:uncert_2D_neutronspectra}
     \end{subfigure}
\caption{\label{fig:2D_neutronspectra} Characterization was performed at a flight-path length of 1 m and angles ranging from 0º to 90º in steps of 5º.}
\end{figure}

The neutron source of the experiment was the reaction \textsuperscript{7}Li(p,n)\textsuperscript{7}Be on a lithium fluoride target of thickness of 14.32~$\mathrm{\mu}$m, held by an aluminum backing. Protons were accelerated up to 2.10 $\pm$ 0.01~MeV. Analysis of the detected neutrons showed energies up to 360~keV. To our knowledge, no experimental characterization of the  energetic-angular profile of emission from the \textsuperscript{7}Li(p,n)\textsuperscript{7}Be reaction at 2.10 MeV has ever been done, so such characterization of the emission from the target was made first. This characterization is of great importance to account for experimental effects and to be used as reference. The reconstructed emission spectra is shown in Figure \ref{fig:2D_neutronspectra}. It was verified that all the experimental TOF spectra that were measured in the laboratory are properly reproduced in our Monte Carlo simulations within uncertainties, which is an important condition to ensure that later calculations (especially those related to multi-scattering) derived from those simulations are valid.

\subsection{Time-of-flight experimental histograms}
Time-of-flight (TOF) is calculated as the difference in timestamp of registered events in the \textsuperscript{6}Li-glass and the timestamp of the Pick-up signal. The capacitive beam pick-up is located just before the neutron-producing target and produces a fast bipolar signal when a beam bunch crosses it. The time reference, i.e., where TOF is equal to zero, is calculated with Monte Carlo simulations in which a \textsuperscript{6}Li-glass with the same dimensions as the experimental one is placed at the same Flight path, by matching the experimentally observed $\gamma$-flash with the simulated $\gamma$-flash.

\begin{figure}
     \centering
     \begin{subfigure}[b]{0.45\textwidth}
         \centering
         \includegraphics[width=\textwidth]{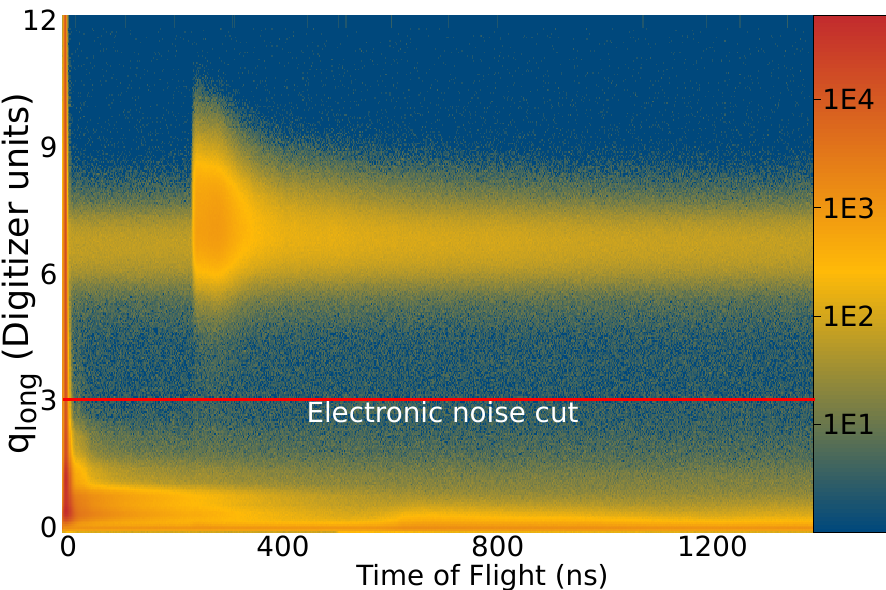}
         \caption{2D-plot of all the registered events during the sample out measurements. There is a clear differentiation between neutrons (around q\textsubscript{long}$\approx$ 7) and electronic noise (mainly below q\textsubscript{long}$\approx$ 3).}
         \label{fig:2D_electronic_noise}
     \end{subfigure}
     \hfill
     \begin{subfigure}[b]{0.46\textwidth}
         \centering
         \includegraphics[width=\textwidth]{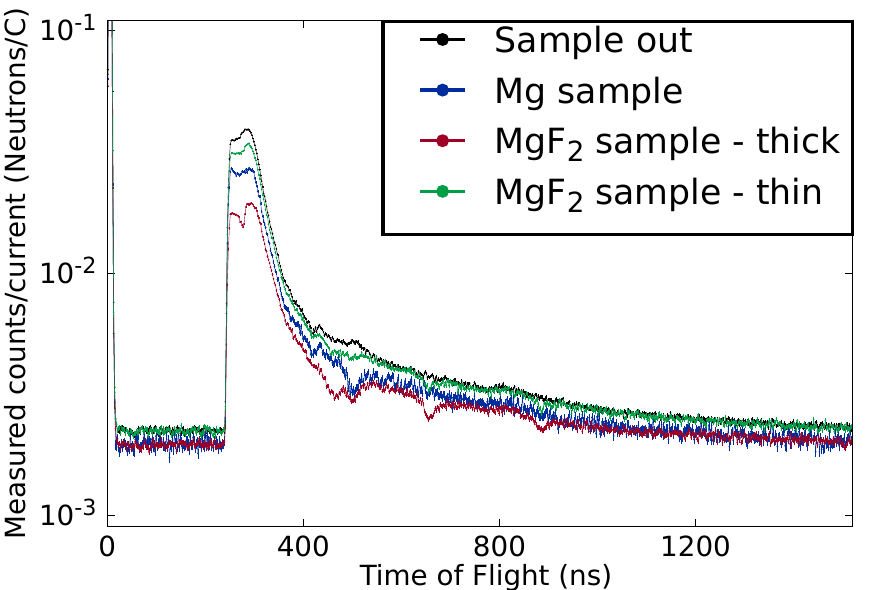}
         \caption{Measured counts normalized by current of the four measurements. Obtained by projecting plots like the on in Figure \ref{fig:2D_electronic_noise} after electronic noise cuts. The background from previous pulses can be clearly seen before the neutron edge and the $\gamma$-flash.}
         \label{fig:tofs_background}
     \end{subfigure}
     \hfill
     \begin{subfigure}[b]{0.46\textwidth}
         \centering
         \includegraphics[width=\textwidth]{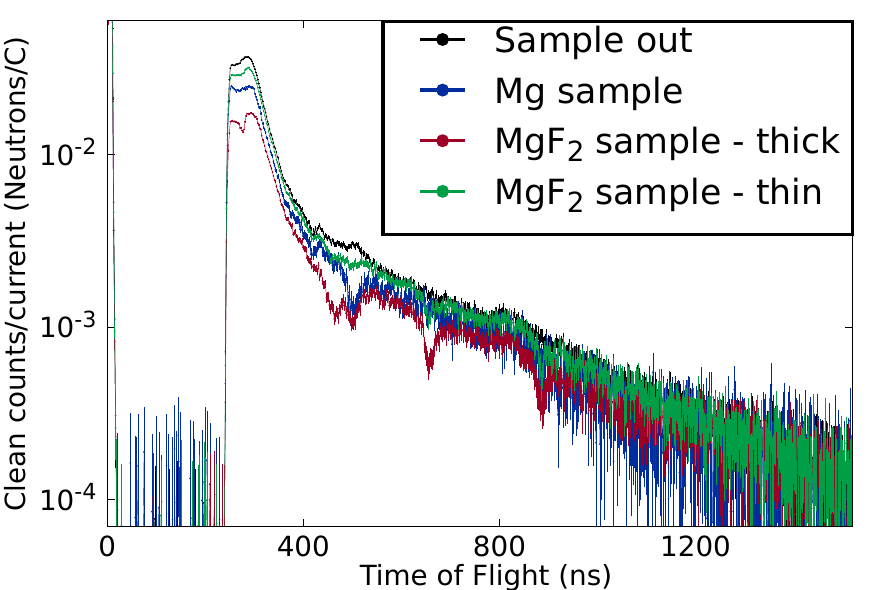}
         \caption{Clean counts obtained by subtraction of the background.}
         \label{fig:tofs_corrected}
     \end{subfigure}
        \caption{}
        \label{fig:three graphs}
\end{figure}

TOF spectra are shown in Figs. \ref{fig:2D_electronic_noise}, \ref{fig:tofs_background} and \ref{fig:tofs_corrected}. Figure \ref{fig:2D_electronic_noise} is a 2D-plot of Time of Flight vs. q\textsubscript{long}, a parameter proportional to the area of the waveform of the detected event. Figure \ref{fig:tofs_background} is a projection of \ref{fig:2D_electronic_noise}, and \ref{fig:tofs_corrected} shows clean and normalized counts, obtained after the following corrections: 
\begin{itemize}
    \item Firstly, electronic noise can be separated from real counts thanks to cuts on q\textsubscript{long}, with corrections to account for lost detections below this cut. A constant threshold cut (see Figure \ref{fig:2D_electronic_noise}) effectively filters out most electronic noise, with minimal loss of real signal counts. An estimate of how many real counts are lost due to the cut is done by fitting the neutron peak in q\textsubscript{long}: very small corrections, around 0.001\% (of all counts), are found for all fits.
    \item No dead-time corrections were applied since the count rate was very low: about 0.0001 count per pulse of protons. 
    \item A background of neutrons is always present, due to the stacking of slow neutrons from past pulses with new ones (see Figure \ref{fig:tofs_background}). Every 1600~ns a new bunch of protons impinges on the LiF target. This stacking has a contribution from each past pulse, and the older the pulse, the lower its contribution. This background contribution from past pulses, B(T), is modeled in this work\footnote{For a step-by-step calculation of B(T) consult the Appendix.} as 
\end{itemize}

The TOF spectra is finally obtained by the following formula:
    \begin{eqnarray}
    C_{C}(T) &= C_{M}(T) - B(T) \nonumber \\
    &=  C_{M}(T) - \Big( \frac{C\cdot e^{-\alpha\cdot T}}{e^{\tau\cdot\alpha}-1} + C_{0}\Big),
    \end{eqnarray}\label{eq:background}

    where
    \begin{enumerate}
        \item $C_{C}$ stands for clean counts.
        \item $T$ is Time of Flight.        
        \item $C_{M}$ stands for experimentally measured counts.        
        \item $\tau$ is the repetition period of a single pulse (1600~ns in our case).
        \item $\alpha$, $C$ and $C_{0}$ are parameters fitted to reproduce the tail of $C_{M}$, with  $C$ and $C_{0}$ measured in the same units as $C_{C}$ and $C_{M}$ and $\alpha$ measured in ns$^{-1}$.
    \end{enumerate}

\subsection{Multi-scattering corrections}
\begin{figure}[]
\includegraphics[width=0.5\textwidth]{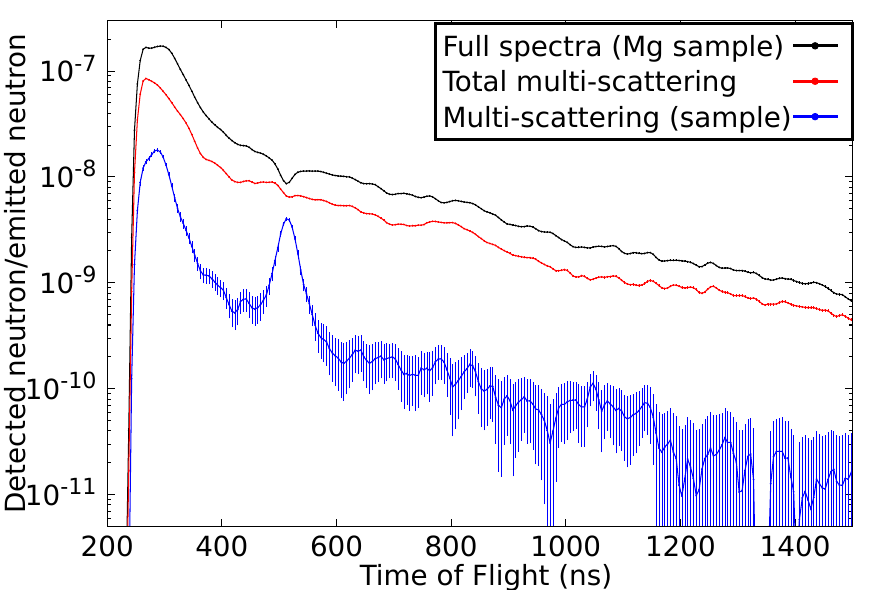}
\caption{\label{fig:multi-scattering} Multi-scattering analysis of the experiment, performed by tallying in function of the number of collisions in MCNP simulations.}
\end{figure}

The most significant correction comes from multi-scattering events. Figure \ref{fig:multi-scattering} shows in black line the expected spectra in the \textsuperscript{6}Li-glass detector when the Mg sample is simulated in MCNP6.3, using the neutron emission spectra from Figure \ref{fig:2D_neutronspectra}. By selectively filtering by number of collisions that a particle has undergone to, before being detected, we are able to calculate the multi-scattering contribution. The total contribution of neutrons that underwent at least one scattering event (whether inside the sample or inside other element of the set up, e.g. floor or walls) is given in the red line, while the blue line shows only the contribution of neutrons that underwent at least one scattering event inside the sample. Two key facts are found: first, about 50\% of detected neutrons are not directly transmitted and absorbed but instead scattered somewhere (even inside the detector itself) before being absorbed; second, that the scattering inside the sample is in general about one order of magnitude smaller than scattering inside other elements of the set up, except near resonances (for instance, around 500~ns) where the scattering cross-section is much higher and scattering events therefore occur more frequently inside the sample. This is particularly caused by the broad angular distribution of the \textsuperscript{7}Li(p,n) at 2.1~MeV. 

The main consequence of these facts is that usual self-shielding and multi-scattering corrections are insufficient to obtain an acceptable cross-section given our experimental conditions. The SAMMY code \cite{SAMMY_guide} includes theoretical models to correct self-shielding, multi-scattering and border effects that take place inside the sample, but those would be insufficient for our case, given that the biggest part of alteration in the cross-section from multi-scattering comes from multi-scattering that occurred outside the sample. Therefore a different approach was taken: the multi-scattering contribution is calculated with simulations and then subtracted from the experimental data. This is why an accurate knowledge of the neutron source was paramount: if the multi-scattering corrections calculated with a certain neutron source are to be trusted, that neutron source must first faithfully reproduce the total measured spectra at all angles.

However, a very subtle question remains: to calculate the multi-scattering correction from Figure \ref{fig:multi-scattering}, cross-sections from ENDF-VIII.1 evaluations for \textsuperscript{19}F, \textsuperscript{24}Mg, \textsuperscript{25}Mg and \textsuperscript{26}Mg isotopes were used in MCNP6.3; but, the cross-section of those isotopes is precisely what we are aiming to calculate. Thus, corrections obtained using ENDF (or any other) libraries would not match the real multi-scattering contribution of those experimentally measured. Such problem is overcome by repeatedly iterating the process until the convergence of cross-sections is reached. The iterative algorithm, represented in the diagram of Figure \ref{fig:algorithm}, is:

\begin{figure}[]
\includegraphics[width=0.5\textwidth]{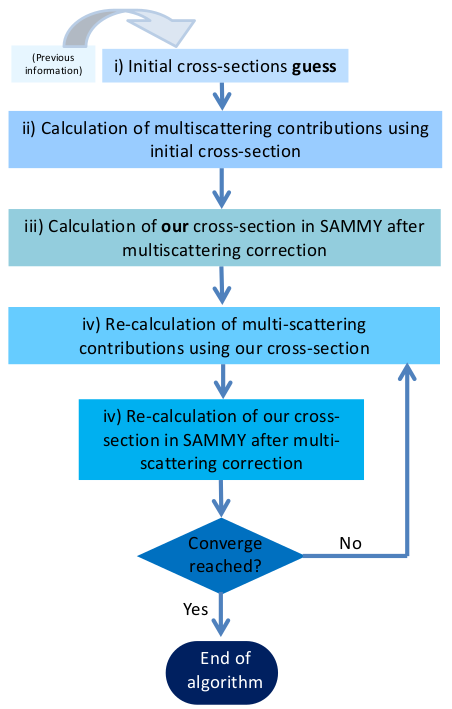}
\caption{\label{fig:algorithm} Iterative algorithm for the calculation of the cross-section.}
\end{figure}

\begin{enumerate}
    \item An initial multi-scattering correction is computed using previous information available (which was the ENDF-VIII.1 evaluation in our case).
    \item Subtraction of those corrections and R-matrix analysis with SAMMY code to obtain our own first cross-section.
    \item Incorporation of our own cross-section into MCNP to recalculate again the multi-scattering correction. The NJOY code \cite{MacFarlane2016NJOY} can take the resonance parameters and generate ACE files which will be read by MCNP. Outside the energy range treated in this work, no modifications were done.
    \item Subtraction of the new corrections and repetition of R-matrix analysis with SAMMY to obtain a new cross-section.
    \item Iteration of points 3 and 4 until the cross-section obtained from SAMMY fitting matches within uncertainty with the cross-section that was last introduced in MCNP code.
\end{enumerate}

\begin{figure*}[]
\includegraphics[width=1.0\textwidth]{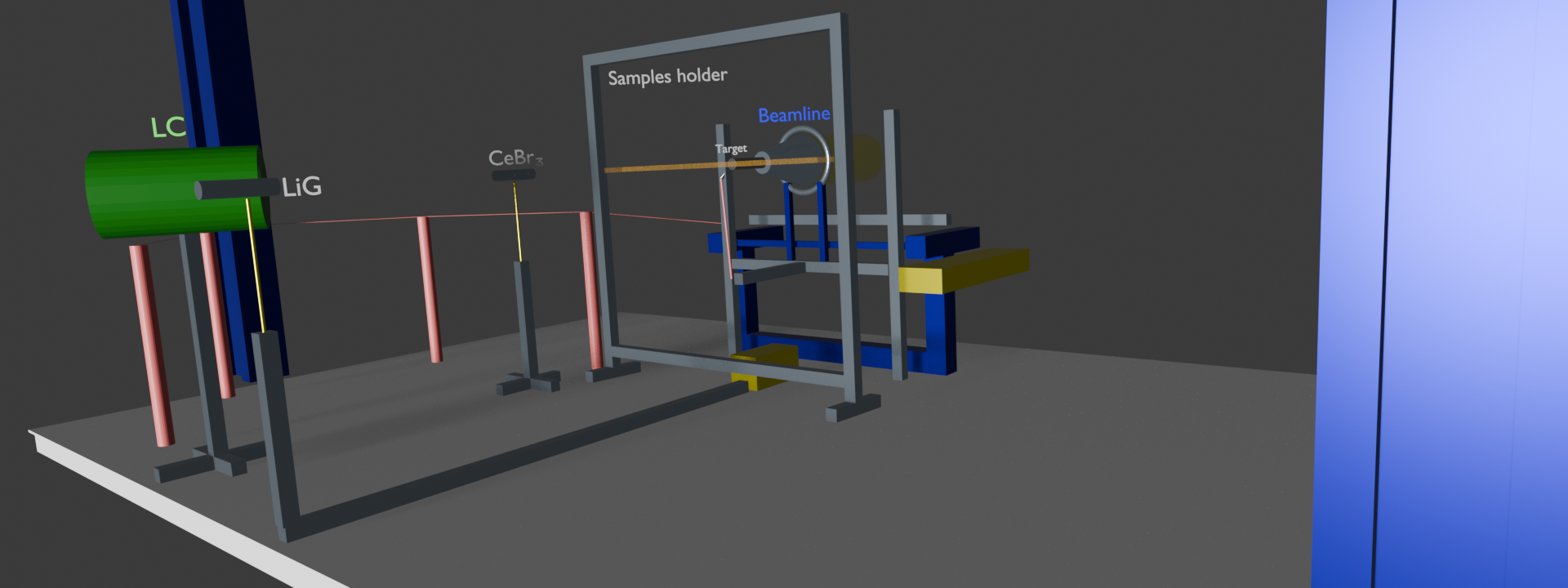}
\caption{\label{fig:simulated_chamber} 3D render of the experimental hall. The elements included in the simulation are those that contribute to the scattering correction, either due to their proximity to the neutron source or the \textsuperscript{6}Li-glass detector, or due to their large mass and extension: beams on the floor, columns, beamline structures, cables, air and the other detectors with their holders.}
\end{figure*}

This method ensures very refined multi-scattering corrections, especially suitable for our experimental set up. The method has several technical requirements to be considered: in the first place, the necessity of a characterization of the neutron source (Figure \ref{fig:2D_neutronspectra}) able to reproduce the measured spectra; secondly, a very detailed geometry for multi-scattering simulations reproducing the experimental hall (Figure \ref{fig:simulated_chamber}) must be included; and finally, previous knowledge of the cross-section to be measured is not mandatory but it notably speeds up the process.

\section{Data analysis}
\subsection{TOF to Energy conversion}
Conversion from Time of Flight to Energy is made  by application of a Resolution Function (RF) calculated using MCNP \cite{MCNP_user_manual}. RF simulations include only the neutron source and the crystal of the detector filled with void, without the rest of the experimental set up, whose effects are already taken into account by subtraction of multi-scattering correction. Therefore, a very narrow distribution is obtained, which mainly accounts for the extension of the source, the extension of the detecting crystal, and the time width of the pulse. The flight path length in the simulations (and therefore, the Resolution Function) slightly varies from sample to sample and was measured using a laser meter with an accuracy of $\pm 0.01$~cm, uncertainty well below that of the time bins resolution that was used in this experiment, 2~ns (a shift in 0.01~cm translates into a change of 0.1~ns at the higher energy of the measured range, 0.6~ns at the lowest).

Conversion to neutron energy is obtained from Time of Flight histograms (after multi-scattering being corrected) by solving

    \begin{eqnarray}
    \overline{T} = \widehat{RF} \cdot \overline{E},
    \end{eqnarray}\label{eq:RF}

where $\overline{T}$ and $\overline{E}$ are vectors associated to time and energy histogram, respectively, and $\widehat{RF}$ is the matrix associated to the RF. Inversion of the matrix is done by using the Expectation-Maximisation (EM) method described in \cite{Tain2007}.

\subsection{Uncertainties}
A conservative approach is taken with respect to uncertainties propagation. 

As for statistical uncertainties, raw measured spectra range from 0.4\% at the neutron edge (around 350~keV) to 2\% at the tail (around 9~keV). Background uncertainty is around 0.3\%. As for systematic uncertainties, those include the determination of the masses of the samples (0.001\%), the flight path (0.05\%) and the current (0.1\%). Lastly, multi-scattering correction contains a systematic contribution (since masses and flight paths are also simulated) and a statistical one (since it is calculated with Monte Carlo methods).

After all corrections, total uncertainty ranges from 1\% at the neutron edge to 9\%, and up to 6\% in the observed resonances. Those final uncertainties are included and propagated via SAMMY to the uncertainty of the varied resonance parameters.

\subsection{R-matrix analysis}
\begin{figure*}[]
\includegraphics[width=1\textwidth]{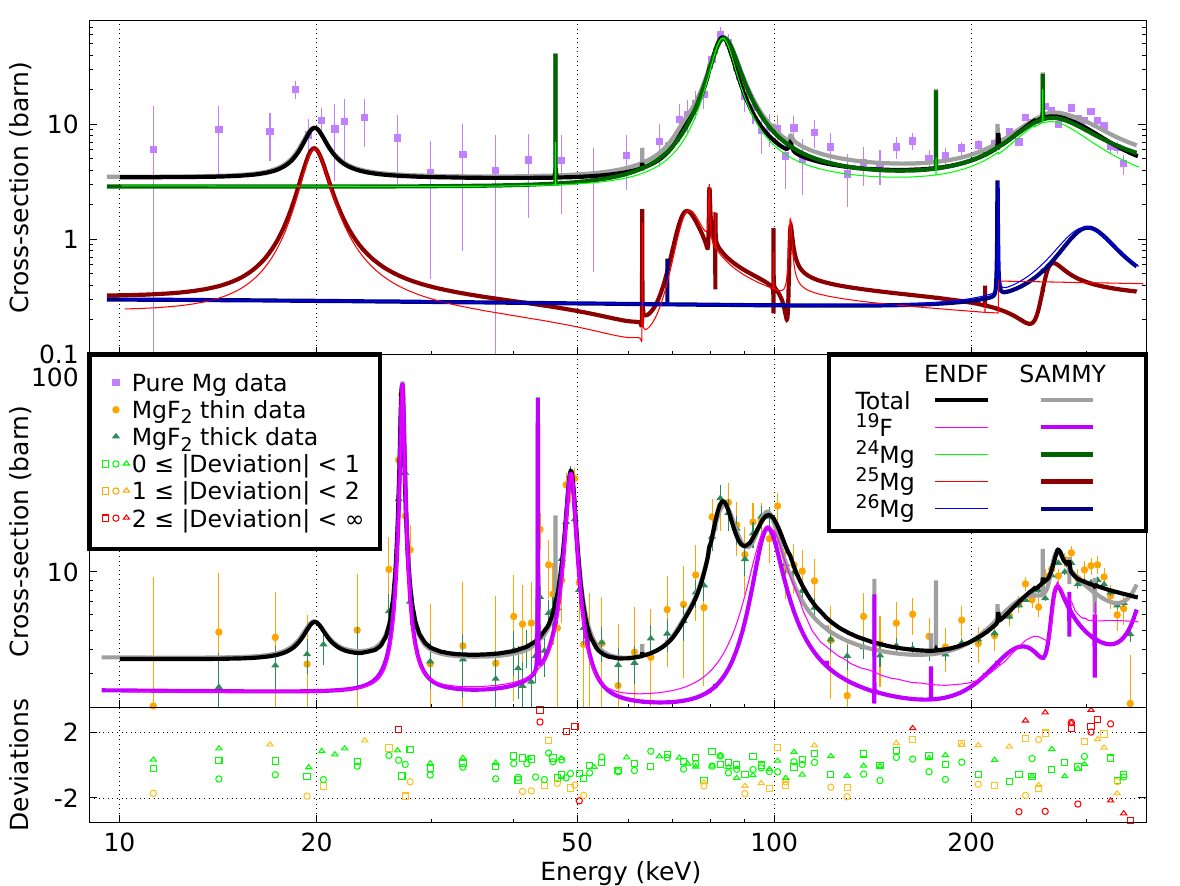}
\caption{\label{fig:final_xs} Converged cross-sections with experimental data and each isotope contribution. The binning used is 50 or 100 bins per decade (bpd) for resonances and 10, 20 or 40~bpd for valleys between resonances. In the lower part of the plot the deviations of each experimental data set with respect to the yield calculated by SAMMY are shown, indicating general good agreement between experiment and calculations.}
\end{figure*}

Cross-section data is fitted with the SAMMY code, where optimization of the resonance parameters (partial widths $\Gamma_c$ of each channel $c$ and resonance energies $E_0$) is performed with R-matrix analysis. Then, a calculation of the theoretical cross-sections is done with the optimal parameters. Atomic radii given by $R=1.23\cdot A^{1/3}$~fm (A being the mass number of the nuclei) have been used during the fitting procedure. Doppler broadening effects are included in the analysis.

The first step involves fitting the parameters associated with magnesium isotopes using data from the pure magnesium sample. These parameters are then kept fixed during posterior analysis of the MgF\textsubscript{2} data to extract the fluorine parameters.

For a given iteration $n$, we define the convergence criterion as $n: \varepsilon^{(n)}(E) \leq 1.0$ $\forall$ E, with $\varepsilon^{(n)}$ defined as

\begin{eqnarray}
    \varepsilon^{(n)}(E) = \frac{\sigma^{(n)}(E) - \sigma^{(n-1)}(E)}{\Delta^{(n)}(E)},
\end{eqnarray}\label{eq:convergence}

where $\sigma^{(n)}(E)$ is cross-section after iteration $n$, and $\Delta(E)$ is the uncertainty (calculated by SAMMY) of $\sigma(E)$. The convergence criterion is equivalent to imposing that the final cross-section at every evaluated energy matches within at least one standard deviation from the cross-section of the previous iteration. As already mentioned, starting from ENDF cross-section, very similar to the cross-section that we aim to calculate, speeds up the process and convergence is quickly reached: five iterations were needed. 

\begin{table*}[]
\caption{Converged resonance parameters drawn from the pure Mg sample: energy of the resonance $E_{R}$, partial widths of elastic neutron scattering channel $\Gamma_{n}$ and width of the gamma channel $\Gamma_{\gamma}$. Values from previous works are also provided: ENDF-VIII.1 evaluation \cite{ENDF_evaluation}, Massimi \textit{et al.} \cite{massimi2017, massimi2012} (2017, 2012), ATLAS of neutron resonances \cite{ATLAS_Mughabghab}, Newson \textit{et al.} \cite{newson1961} and Singh \textit{et al.} \cite{singh1974_Mg}.}\label{tab:resonance_parameters_Mg}
\resizebox{1\textwidth}{!}{\begin{tabular}{|c|c|c|c|ccccccc|}
\hline
& $\mathbf{E_{R}}$ & & $\mathbf{\Gamma_{\gamma}}$ & \multicolumn{7}{c|}{$\mathbf{\Gamma_n}$ \textbf{(keV)}}  \\
\cline{5-11}
\multirow{-2}{*}{ \textbf{Isotope} } & \textbf{(keV)} & \multirow{-2}{*}{$\mathbf{J^{\pi}}$} & \textbf{(eV)}
  & \multicolumn{1}{c|}{This work}
  & \multicolumn{1}{c|}{\cite{ENDF_evaluation}}
  & \multicolumn{1}{c|}{\cite{massimi2012}}
  & \multicolumn{1}{c|}{\cite{massimi2017}}
  & \multicolumn{1}{c|}{\cite{ATLAS_Mughabghab}}
  & \multicolumn{1}{c|}{\cite{newson1961}}
  & \multicolumn{1}{c|}{\cite{singh1974_Mg}} \\
\hline
\multirow{2}{*}{\textsuperscript{24}Mg} & \multicolumn{1}{|c|}{83.940}     & \multicolumn{1}{c|}{$\tfrac{3}{2}^{-}$} & \multicolumn{1}{c|}{4.1}
     & \multicolumn{1}{c|}{9.2 (0.6)}
     & \multicolumn{1}{c|}{9}
     & \multicolumn{1}{c|}{7.607(0.004)}
     & \multicolumn{1}{c|}{- }
     & \multicolumn{1}{c|}{7.7(0.5)}
     & \multicolumn{1}{c|}{7.8(0.5)}
     & 7.6(0.5)                       \\
 & \multicolumn{1}{|c|}{267.480}    & \multicolumn{1}{c|}{$\tfrac{1}{2}^{-}$} & \multicolumn{1}{c|}{7}
     & \multicolumn{1}{c|}{92(4)}
     & \multicolumn{1}{c|}{115}
     & \multicolumn{1}{c|}{83.270(0.020)}
     & \multicolumn{1}{c|}{- }
     & \multicolumn{1}{c|}{79.0(0.5)}
     & \multicolumn{1}{c|}{75(15)}
     & 90(5)                                       \\
\textsuperscript{25}Mg & \multicolumn{1}{|c|}{72.66}      & \multicolumn{1}{c|}{$2^{+}$}            & \multicolumn{1}{c|}{2.5}
     & \multicolumn{1}{c|}{7.5(1.5)}
     & \multicolumn{1}{c|}{7.6}
     & \multicolumn{1}{c|}{5.08(0.08)}
     & \multicolumn{1}{c|}{5.31(0.05)}
     & \multicolumn{1}{c|}{4.6(1.1)}
     & \multicolumn{1}{c|}{12(3)}
     & 7.6(0.5)                                  \\
\textsuperscript{26}Mg & \multicolumn{1}{|c|}{302.340}    & \multicolumn{1}{c|}{$\tfrac{1}{2}^{-}$} & \multicolumn{1}{c|}{6.3}
     & \multicolumn{1}{c|}{66(13)}
     & \multicolumn{1}{c|}{90}
     & \multicolumn{1}{c|}{61.2(0.2)}
     & \multicolumn{1}{c|}{-}
     & \multicolumn{1}{c|}{66.92(0.17)}
     & \multicolumn{1}{c|}{$>$ 75}
     & \multicolumn{1}{c|}{-}             \\
\hline
\end{tabular}}
{ \\}
\caption{Converged resonance parameters of \textsuperscript{19}F, drawn from the MgF\textsubscript{2} samples: energy of the resonance \(E_{R}\), widths of (in)elastic neutron scattering channels \(\Gamma_{n}\) and \(\Gamma_{n'}\) and width of the gamma channel \(\Gamma_{\gamma}\). Values from previous works are also provided: ENDF-VIII.1 evaluation \cite{ENDF_evaluation}, ATLAS of neutron resonances \cite{ATLAS_Mughabghab}, Singh \textit{et al.} \cite{singh1974_F} and Macklin \textit{et al.} \cite{macklin1973}.}\label{tab:resonance_parameters_F}
\resizebox{1\textwidth}{!}{\begin{tabular}{|c|c|c|c|cccc|cc|}
\hline
 & $\mathbf{E_{R}}$ & & $\mathbf{\Gamma_{\gamma}}$ & \multicolumn{4}{c|}{$\mathbf{\Gamma_n}$ \textbf{(keV)}} & \multicolumn{2}{c|}{$\mathbf{\Gamma_{n'}}$ \textbf{(keV)}} \\
\cline{5-10}
\multirow{-2}{*}{ \textbf{Isotope} }  & \textbf{(keV)} & \multirow{-2}{*}{ {$\mathbf{J^{\pi}}$} } & \textbf{(keV)}
  & \multicolumn{1}{c|}{This work}
  & \multicolumn{1}{c|}{\cite{singh1974_F}}
  & \multicolumn{1}{c|}{\cite{macklin1973}}
  & \multicolumn{1}{c|}{\cite{ATLAS_Mughabghab}}
  & \multicolumn{1}{c|}{This work}
  & \multicolumn{1}{c|}{\cite{ATLAS_Mughabghab}} \\
\hline
\multirow{5}{*}{\textsuperscript{19}F}
  & 27.052     & \(2^{-}\)   & 1.16  
    & 0.360(0.015) & 0.325(0.020) & –           & 0.356(0.004) & –           & –           \\
  & 48.931     & \(1^{-}\)   & 1.13  
    & 1.73(0.06)   & 1.67(0.10)   & 1.9(0.5)    & 1.80(0.02)  & –           & –           \\
  & 97.946     & \(1^{-}\)   & 1.5   
    & 10.2(0.4)    & 14.5(0.8)    & 13.5(1.5)   & 13.96(0.10) & –           & –           \\
  & 261.220    & \(2^{-}\)   & 21.0  
    & 42.1(1.2)    & –            & –           & 32          & 88(3)       & 61          \\
  & 268.110    & \(1^{+}\)   & 3.1   
    & 14.7(0.5)    & –            & 10(2)       & 8(3)        & 2.27(0.11)  & 2.2         \\
\hline
\end{tabular}}
\end{table*}

Figure \ref{fig:final_xs} includes the cross-sections obtained after fitting of the last iteration, experimental cross-section values and contribution of each isotope, for both cross-sections of each Mg and MgF\textsubscript{2} samples. Five iterations were needed before convergence was reached.

\begin{figure}[]
\includegraphics[width=0.47\textwidth]{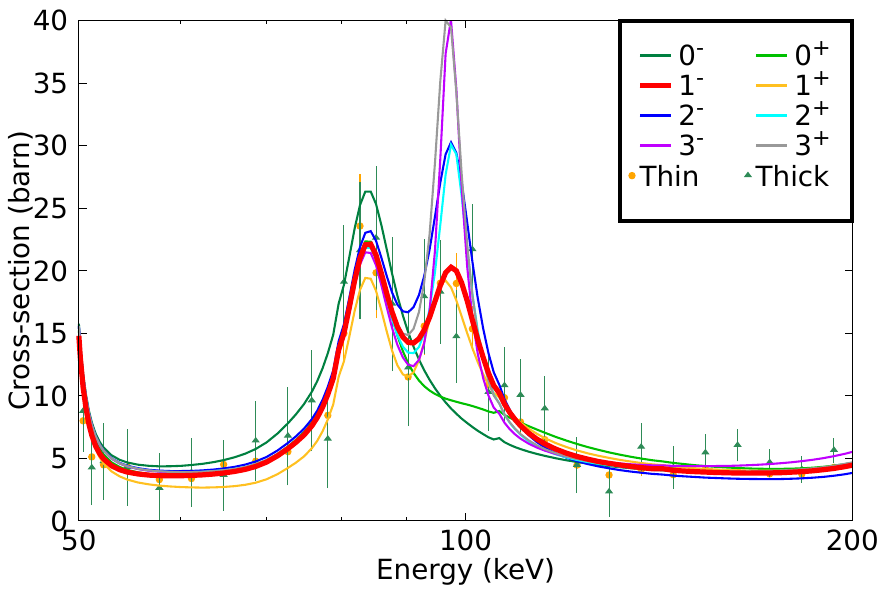}
\caption{\label{fig:spins_reso_19F} Different cross-sections calculated by SAMMY after considering several spin-parity possibilities for the resonance at 97.95 keV of \textsuperscript{19}F. The optimal value, which reduced overall sum of residuals, is $J^{\pi} = (1^{-})$.}
\end{figure}

Tables \ref{tab:resonance_parameters_Mg}, \ref{tab:resonance_parameters_F} contain all the fitted parameters and the uncertainties offered by SAMMY. Parameters previously calculated by other authors are also included for comparison. Given the experimental resolution and uncertainties, the fitting process is only sensible to resonances included in the table, so parameters from \cite{ATLAS_Mughabghab} were used for the rest of resonances. According to the same source, parity and spin of resonance at 97~keV of \textsuperscript{19}F is unknown. Several possibilities were considered in the fitting and $J^{\pi} = 1^{-}$ offered the best fit after optimization of parameters, which agrees with what \cite{singh1974_F} proposed (see Figure \ref{fig:spins_reso_19F}).

As for magnesium parameters, all resonance parameters show a general agreement with previous works. The natural abundance of \textsuperscript{25}Mg is approximately seven times lower than that of \textsuperscript{24}Mg, resulting in significantly reduced sensitivity to the contribution from \textsuperscript{25}Mg in our work. This is reflected in the larger relative uncertainty associated with the resonance parameters of \textsuperscript{25}Mg and \textsuperscript{26}Mg. Particularly, our values are several standard deviations away from those reported by Massimi \textit{et al.}, however the results from Massimi \textit{et al.}  (\cite{massimi2012} in 2012 and \cite{massimi2017} in 2017) also happen to be notably distant from each other, which would suggest an underestimation of their real uncertainty. The largest difference happens for \textsuperscript{25}Mg from Newson \textit{et al.}, not only when compared to our results but when compared to the rest of authors.

Regarding fluorine, a very good agreement is shown for all previously reported resonance parameters. Such comparison is not possible for the inelastic channel since no previous work measuring resonance widths of inelastic reaction has been found.

\subsection{Inelastic cross-section calculation}
\begin{figure}[]
\includegraphics[width=0.47\textwidth]{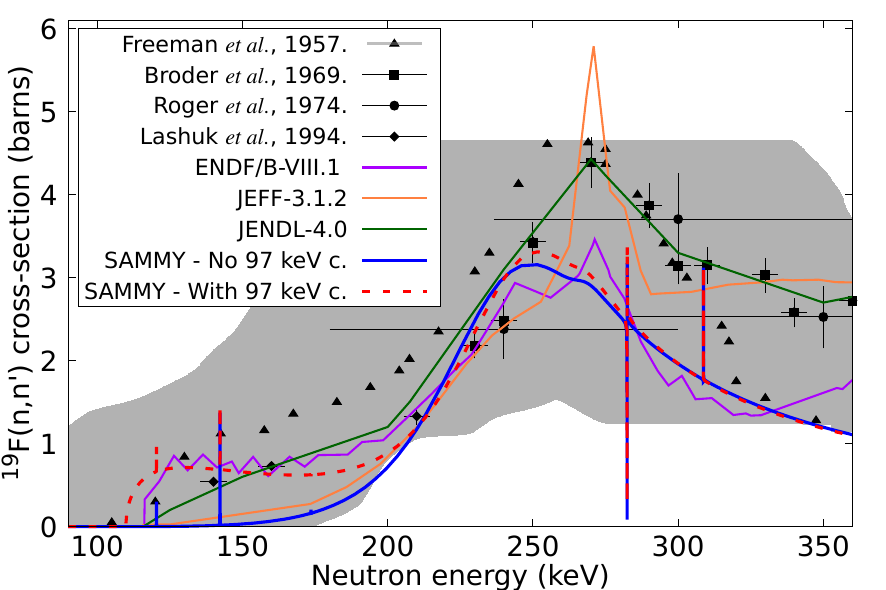}
\caption{\label{fig:inelastic_xs} Inelastic cross-sections calculated by SAMMY considering contribution or not from the resonance 97.95~keV: blue line without contribution, red line with it. In thinner continuous line, three evaluations from ENDF/B-VIII.1, JEFF-3.1.2 and JENDL-4.0 are also provided, while data points correspond to measurements reported in EXFOR.}
\end{figure}

With parameters from Table \ref{tab:resonance_parameters_F}, we are able to calculate the contribution to the total cross-section of each partial reaction: elastic, capture and inelastic. For the reasons given at Section \ref{sec:applications}, the inelastic channel is of particular interest. Opening at about 110~keV, there are a few measurements of the inelastic cross-section, but they are rather focused on higher energies, above 350~keV (upper energy limit of this work), where the inelastic contribution to the total cross-section is larger. Therefore, very scarce points exist between 100 and 300~keV with extremely poor energy resolution in general (which accounts for the large horizontal uncertainties and the grey band in Figure \ref{fig:inelastic_xs}). Moreover, strong disagreements exist between different nuclear evaluations ENDF/B-VIII.1 \cite{ENDF_inelastic}, JEFF-3.1.2 \cite{JEFF_3_1} or JENDL-4.0 \cite{JENDL_4_1}. All these evaluations and measured data are shown in Figure \ref{fig:inelastic_xs}, where the inelastic cross-section calculated with SAMMY using the parameters from Table \ref{tab:resonance_parameters_F} is also displayed in blue.

According to \cite{ATLAS_Mughabghab}, the value of partial width $\Gamma_{n'}$ is 0 for the resonance at 97.95~keV, and in the same way, inelastic contribution is neglected in \cite{singh1974_F}. Following that, in the previous analysis of this work we kept that value as 0. However, a non-zero value could also be plausible. An alternative fitting, obtaining values of $\Gamma_{n}=$ 7.57~keV and $\Gamma_{n'}=$ 105.84~keV, is provided in red dashed line in Figure \ref{fig:inelastic_xs} too, which better matches data from Lashuk \textit{et al.} .

ENDF and JEFF evaluations strongly differ from 100 to 300~keV, especially at the lower range: ENDF predicts higher values below 200~keV, while JEFF predicts higher values above 200~keV. In the range from 110 to 200~keV, where inelastic cross-section contributes the least and the statistics of our measurement is lower, the value of the inelastic cross-section is almost entirely dominated by $\Gamma_{n'}$ of the resonance at 97.95~keV, which we have kept at 0, thus translating into values closer to JEFF than to ENDF.  However, from above 300~keV (where our statistics is better), it is the contribution from $\Gamma_{n'}$ of resonances at 261.22 and 268.11~keV the one that dominates, and according to the fitting we obtained here, ENDF evaluation is in better agreement with our results.

\section{Conclusions}
Total cross-sections have been calculated for isotopes \textsuperscript{19}F, \textsuperscript{24}Mg, \textsuperscript{25}Mg and \textsuperscript{26}Mg from experimental transmission measurements. A refined iterative multiple-scattering correction has been applied and detailed in this work, as well as the requirements and limitations. R-matrix analysis has been performed using the SAMMY code to determine optimized resonances parameters of observed resonances, and allowed for endorsement of $1^{-}$ as the spin and parity of the resonance around 97~keV of fluorine. Parameters of the inelastic channels of fluorine have also been provided for the first time, which could be very useful for future evaluations of the inelastic cross-section. All these data will also be of great help in the development of Beam Shaping Assemblies for BNCT treatment using MgF\textsubscript{2} as moderator. Finally, the developed iterative algorithm is also a very useful result that can be applied in the calculation of any total cross-section in which scattering constitutes such an important contribution. 

\section{Acknowledgements}
We acknowledge support from the EURATOM coordination and support project "Accelerator and Research reactor Infrastructures for Education and Learning (ARIEL) for funding the experiment.

We acknowledge the help received from the JRC-Geel Workshop for the manipulation and re-shaping of the samples to fit the necessities of the experiment.

M.A.M. acknowledges support from the ARIEL program for funding his stay at the Joint Research Centre - Geel, as well as the Spanish Ministry of Science, Innovation and Universities under the FPU Grant No. FPU21/02919. 

\section*{Appendix}
An estimation of the background is made as follows. A first assumption of decay is made by observing the tail of Time of Flight histograms: the amount of neutrons produced from one single pulse decays following an exponential law in time, added to a small constant contribution $C_0$. Therefore, the observed spectrum of emitted neutrons per single proton pulse, $C_1(T)$, follows the equation,

\begin{equation}
    C_1(T) = C\cdot e^{-\alpha\cdot T} + C_{0} \quad ,
\end{equation}

where the second pulse contains a contribution from the first one according to

\begin{equation}
    C_2(T) = C\cdot (e^{-\alpha\cdot T} + e^{-\alpha\cdot (T+\tau )}) + C_{0} \quad .
\end{equation}

Thus, for the $N$-th pulse a sum of all previous pulses is taken into account:

\begin{eqnarray}
    C_N(T) = C\cdot \sum_{n=1}^{N}e^{-\alpha\cdot (T+\tau\cdot(n-1))} + C_{0} \nonumber \\
    =  C e^{-\alpha T}\Big(\frac{1-e^{-\tau\cdot\alpha\cdot N}}{1-e^{-\tau\cdot\alpha}}\Big) + C_{0} \nonumber \\
    \approx  C \Big(\frac{e^{-\alpha\cdot T}}{1-e^{-\tau\cdot \alpha}}\Big) + C_{0} \quad ,
\end{eqnarray}

where $\alpha$ is measured in ns${}^{-1}$. Since the pulse-repetition rate is at least 625~kHz, $N$ is very large, which justifies the approximation $e^{-\tau\cdot \alpha N} \approx 0$.
The function is fitted to the final range of the experimental histograms (from 1200~ns onward) to estimate values of $C_{0}$ and $\alpha$. What we are really interested in is the production from one single pulse, so the correction that must be subtracted, B(T), is calculated as

    \begin{eqnarray}
    B(T) = C_N(T) - C_1(T) + C_0 \\
=  C e^{-\alpha T}\Big(\frac{1-e^{-\tau \cdot \alpha \cdot N}}{1-e^{-\tau\cdot \alpha}}\Big) - C\cdot e^{-\alpha\cdot T} + C_0 \nonumber \\
\approx C\frac{e^{-\alpha T}}{e^{\tau\cdot\alpha} - 1} + C_{0}.
    \end{eqnarray}

A rougher approximation would have been also possible by assuming a simple constant background, since the time dependency quickly fades. For our case, the largest difference is found at the neutron edge with the time-dependent background being 2\% above the constant estimate. That difference is small, though still within our sensitivity, and therefore time dependency is included.

\section*{References}
\bibliographystyle{iopart-num}

\providecommand{\newblock}{}


\end{document}